# Evidence of coexistence of change of caged dynamics at $T_g$ and the dynamic transition at $T_d$ in solvated proteins

S. Capaccioli[a,b][1], K. L. Ngai[a][2], S. Ancherbak[b], A. Paciaroni[c]

[a]*CNR-IPCF, Dipartimento di Fisica, Largo Bruno Pontecorvo 3, I-56127, Pisa, Italy*
[b]*Dipartimento di Fisica, Università di Pisa, Largo Bruno Pontecorvo 3, I-56127, Pisa, Italy*
[c]*Dipartimento di Fisica, Università di Perugia & IOM-CNR, , Via A. Pascoli 1, 06123 Perugia, Italy*

---

**Abstract**

Mössbauer spectroscopy and neutron scattering measurements on proteins embedded in solvents including water and aqueous mixtures have emphasized the observation of the distinctive temperature dependence of the atomic mean square displacements, $<u^2>$, commonly referred to as the dynamic transition at some temperature $T_d$. At low temperatures, $<u^2>$ increases slowly, but it assumes stronger temperature dependence after crossing $T_d$, which depends on the time/frequency resolution of the spectrometer. Various authors have made connection of the dynamics of solvated proteins including the dynamic transition to that of glass-forming substances. Notwithstanding, no connection is made to the similar change of temperature dependence of $<u^2>$ obtained by quasielastic neutron scattering when crossing the glass transition temperature $T_g$, generally observed in inorganic, organic and polymeric glass-formers. Evidences are presented here to show that such change of the temperature dependence of $<u^2>$ from neutron scattering at $T_g$ is present in hydrated or solvated proteins, as well as in the solvent used, unsurprisingly since the latter is just another organic glass-former. If unaware of the existence of such crossover of $<u^2>$ at $T_g$, and if present, it can be mistaken as the dynamic transition at $T_d$ with the ill consequences of underestimating $T_d$ by the lower value $T_g$, and confusing the identification of the origin of the dynamic transition. The $<u^2>$ obtained by neutron scattering at not so low temperatures has contributions from the dissipation of molecules while caged by the anharmonic intermolecular potential at times before dissolution of cages by the onset of the Johari-Goldstein β-relaxation or of the merged α-β relaxation. The universal change of $<u^2>$ at $T_g$ of glass-formers, independent of the spectrometer resolution, had been rationalized by sensitivity to change in volume and entropy of the dissipation of the caged molecules and its contribution to $<u^2>$. The same rationalization applies to hydrated and solvated proteins for the observed change of $<u^2>$ at $T_g$.

---

[1] capacci@df.unipi.it, Tel. +39-0502214537, Fax. +39-0502214333
[2] Ngai@df.unipi.it, Tel. +39-0502214322, Fax. +39-0502214333

# 1. Introduction

Relaxation and diffusion originating from molecular motions in solvated or simply hydrated proteins naturally and ultimately are responsible for the dynamics that give rise to biological functions. In one way or the other, various approaches to understand the dynamics of proteins have been based on exploiting the similarity of the properties observed experimentally to the dynamics of glass-forming systems [1, 2, 3, 4, 5, 6, 7]. In fact, the very basic glass transition was observed in solvated proteins by calorimetry [7, 8, 9, 10, 11, 12], thermal expansion measurements [13], and Brillouin scattering [14], with the glass transition temperature $T_g$ decreasing on increasing the hydration level and generally falls within the range of 160 K to 200 K, and can be higher if water is totally absent in the solvent such as pure glycerol or the solvent is 20 wt% of water in the disaccharide, sucrose [12, 13]. Besides the relation to glass transition, another general phenomenon exhibited by solvated proteins which has occupied much attention is the so-called dynamic transition (*i.e.*, the anharmonic onset of molecular displacements given by the mean square displacement $<u^2>$ at temperature $T_d$) observed for instance by either Mössbauer spectroscopy [15, 16] or by neutron scattering [7, 17, 18, 19, 20, 21, 22, 23, 24, 25, 26, 27, 28, 29, 30, 31, 32, 33, 34]. The protein dynamic transition temperature $T_d$ depends on the time scale or energy resolution of the spectrometer used. In the 140 ns long time limit of Mössbauer spectroscopy the $T_d$ measured in deoxy-myoglobin crystals is at slightly below 200 K [15]. Neutron scattering measurements with time scale ranging from about 1 ns to 15 ps show the dynamic transition occurring at higher $T_d$ than that from Mössbauer spectroscopy. For fully hydrated myoglobin and lysozyme, $T_d$ from neutron scattering increase from about 210 K at time scale of 1 ns to 250-260 at 15 ps. The explanations of the origin of the dynamic transition vary from one research group to another [6, 7, 28, 34, 35, 36, 37, 38, 39, 40, 41, 42, 43, 44, 45]. A recent experiment on dry and hydrated lysozyme [46] confirmed, by means of different neutron scattering experiments with different instrumental energy resolutions, that the dynamic transition observed at $T_d$ appears when the characteristic system relaxation time intersects resolution time and so it is instrument dependent. The glass transition and the dynamic transition are therefore two different processes in the same solvated or hydrated protein, and $T_d$ is higher than $T_g$, in general.

In ordinary glassformers, the fast dynamics measured as a function of temperature from below to above $T_g$ and expressed in terms of $<u^2(T)>$ exhibit a universal behaviour. As measured by the same neutron scattering technique and spectrometers including IN6, IN13 and IN16 as for dynamic transition in solvated proteins, $<u^2(T)>$ of ordinary glassformers has a weak and approximately linear temperature dependence for $T<T_g$ but it assumes a stronger $T$-dependence after crossing $T_g$ [47]. The sensitivity of the fast process manifesting in the ps to ns range to glass transition is remarkable because glass transition is affected by the structural α-relaxation when, on cooling, its relaxation time becomes too long compared with the experimental time typically of the order of $10^3$ s, and the liquid can no longer maintain equilibrium. The result is vitrification and the transformation into the glassy state. This general property found in inorganic, organic and polymeric glassformers was touted as one of the important aspects in the dynamics of glass transition [47, 48, 49, 50, 51, 52, 53]. At temperatures below and slightly above $T_g$, the fast process observed at times shorter than 1 ns by neutron scattering comes from motion of molecules mutually caged by anharmonic potential. Unlike genuine relaxation process, the loss part of the susceptibility, $\chi''$, from the molecules while caged has no characteristic time and hence its dependence on frequency $\omega$ is a power law, $\chi'' \sim \omega^{-c}$, where $c$ is a small positive number. Dielectric relaxation spectroscopy also has observed this characteristic frequency dependence of the dielectric loss $\varepsilon'' \sim \omega^{-c}$, which is generally referred to as the nearly constant loss (NCL) since $c \ll 1$ [54, 55]. When cages decay with the onset of the intermolecular secondary relaxation or the so-called Johari-Goldstein (JG) β-relaxation, the regime of caged dynamics and the associated NCL are terminated. It has been well established by experiments that the temperature dependence of both the dielectric strength and the relaxation time of the JG β-relaxation changes when crossing $T_g$ [56, 57, 58, 59] in conventional glassformers, in aqueous mixtures [60, 61], and in hydrated proteins [41, 43, 60]. Thus, this property of the JG β-relaxation is transferred to the NCL, and is currently the only explanation that has been offered for the change of $T$-dependence of $<u^2(T)>$ at $T_g$ observed by neutron scattering, dynamics light scattering, and dielectric relaxation [62, 52, 53].

From the similarity of solvated protein dynamics to glass transition, it is natural to inquire whether the change in $T$-dependence of $<u^2(T)>$ on crossing $T_g$ found in ordinary glassformers has been observed or not in solvated or hydrated proteins. This question has not been addressed before, at least correctly for the experimental data of solvated and hydrated proteins, by anyone as far as we know. If indeed the change of $T$-dependence of $<u^2(T)>$ at $T_g$ has been observed in different solvated proteins, then it should be considered as another remarkable property challenging an explanation in addition to the dynamic transition. Since $T_g$ and $T_d$ in some cases are not far apart, if it has been observed, could it be mistaken before as the dynamic transition at $T_d$? We are mindful of the contribution of methyl group rotation to $<u^2(T)>$ at temperatures higher than about 100 K in typical high-energy resolution neutron investigations [3, 63, 34, 31, 64, 65, 66]. The methyl group contribution makes it more difficult to identify the crossover of $<u^2(T)>$ at $T_g$ in cases where $T_g$ is not much higher than 150 K. Notwithstanding, the purpose of this paper is to answer the questions posed above by re-examining published experimental data of solvated and hydrated proteins. An assist is given by the neutron scattering data of the solvents themselves by showing the presence of the change of $T$-dependence of $<u^2(T)>$ at $T_g$ and at a higher temperature $T_d$, in exact analogy to that observed in protein solvated by the same solvent.

## 2. Re-examination of Published Experimental Data

In the following sections we will re-examine some already published results, coming from neutron scattering experiments done on several spectrometers and concerning proteins in different solvents or the solvent alone. We will also present a new set of data for the samples of lysozyme:glycerol and lysozyme:glucose (both at 50:50 relative weight) in an extended temperature range, which will be the starting point to propose a coherent interpretation of the experimental results. The main experimental and sample characteristics are summarized in Table 1.

| **Reviewed Systems** | **Instrument** | **ΔE(μeV)** | **Ref.** |
|---|---|---|---|
| Lysozyme+glycerol(D)+variable D$_2$O | IN13 | 9 | 19 |
| Lysozyme+glycerol(D) | HFBS | 1 | 20 |
| Glycerol (part. Deuter.) | IN13 | 9 | 67 |
| Glycerol | IN10-IN16<br>TOF- Stockholm | 1,<br>200 | 68,<br>81 |
| Glycerol+H$_2$O (0.15 g water/g Glyc) | NEAT | 10÷1000 | 44 |
| Lysozyme+D$_2$O 0.4$h$ | IN16 | 1 | 24 |
| Lysozyme+D$_2$O (variable $h$) | HFBS | 1 | 19 |
| Lysozyme+D$_2$O 0.4$h$ | IN13 | 9 | 69 |
| Lysozyme+Glucose (D)+variable D$_2$O | IN13 | 9 | 69, 70 |
| Glucose+variable D$_2$O | IN13 | 9 | 71, 72 |
| Disaccharides+variable D$_2$O | IN13 | 9 | 73, 74, 75 |
| PM (select. Deut.) +variable D$_2$O | IN16, IN10 | 1 | 76, 77, 78 |
| Myoglobin+D$_2$O 0.35$h$ | IN6 | 90 | 28 |
| C-PC+D$_2$O 0.3$h$ | SPHERES | 0.62 | 28 |

**Table 1**. Experimental data sets reviewed in this paper, in order of appearance. The energy resolution $\Delta E$ (Full width at half maximum FWHM) is approximately linked to the longest measurable characteristic time $\tau_{res}$ by the relationship $\tau_{res}$(ps) ~1316/$\Delta E$(μeV)

*2.1 Lysozyme solvated in glycerol at different contents of water*

These solvated lysozyme systems have been studied by two different groups [19, 20]. Tsai et al. made elastic neutron scattering measurements on dry D-exchanged lysozyme:glycerol (80:20) and (50:50), as well as lysozyme:D$_2$O (70:30), using the neutron high flux backscattering instrument of the National Institute of Standards and Technology. The incident wavelength is at 6.271 Å with an energy resolution of 1 μeV (FWHM) picking up motions faster than about 1 ns. Paciaroni et al. performed elastic neutron scattering studies of lysozyme solvated in glycerol by using the backscattering spectrometer IN13, at the Institut Laue-Langevin, having an energy resolution 9 μeV (FWHM), which makes accessible only motions faster than about 150 ps in a spatial region smaller than ~5 Å. The samples studied include the lysozyme:glycerol (50:50) as Tsai et al., and this lysozyme:glycerol mixture hydrated to the levels of 0.1$h$, 0.2$h$, 0.35$h$, 0.42$h$, and 0.83$h$, where $h$ stands for grams of water/grams of lysozyme. There are neutron scattering data on the solvents alone without lysozyme. Since it has been established by experiments by several groups that the dynamics of solvated proteins are coupled to the solvent, naturally the data of neat glycerol obtained by means of IN13 by Wuttke et al. [67] and by means of IN10 and IN16 by K. Niss et al. [68], and the data of glycerol-15% water by Mezei et al. [44] on the NEAT time-of-flight spectrometer are helpful for interpretation of the data of the solvated lysozyme as can be seen from the discussion after the data have been presented.

In Figure 1 we reproduce the data of the total $<u^2(T)>$ of lysozyme in fully deuterated glycerol (50:50) with 0$h$ [20] together with the mean square displacements of partially deuterated bulk glycerol $C_3H_5(OD)_3$, also measured on the spectrometer IN13 of Wuttke et al. [67]. Here, $<u^2(T)>_{tot}$ is the total mean square displacements obtained from the intensity of elastically scattered neutrons $S(Q, E \approx 0)$ as explained in Ref.[20]. The $<u^2(T)>$ of lysozyme-glycerol (50:50) and pure glycerol are strikingly similar. At lower temperatures, $<u^2(T)>$ has nearly linear $T$-dependence but on increasing temperature past $T_g \approx 190$ K of pure glycerol, it changes to a stronger $T$-dependence after crossing 200-210 K. As shown before in Ref.[47], this change of $T$-dependence of $<u^2(T)>$ above $T_g$ is an universal property of glassformers, and thus Fig.1 suggests that the same property is found in lysozyme-glycerol (50:50). The inset of Fig.1 and its relation to the main part of the figure will be discussed at the end of this subsection.

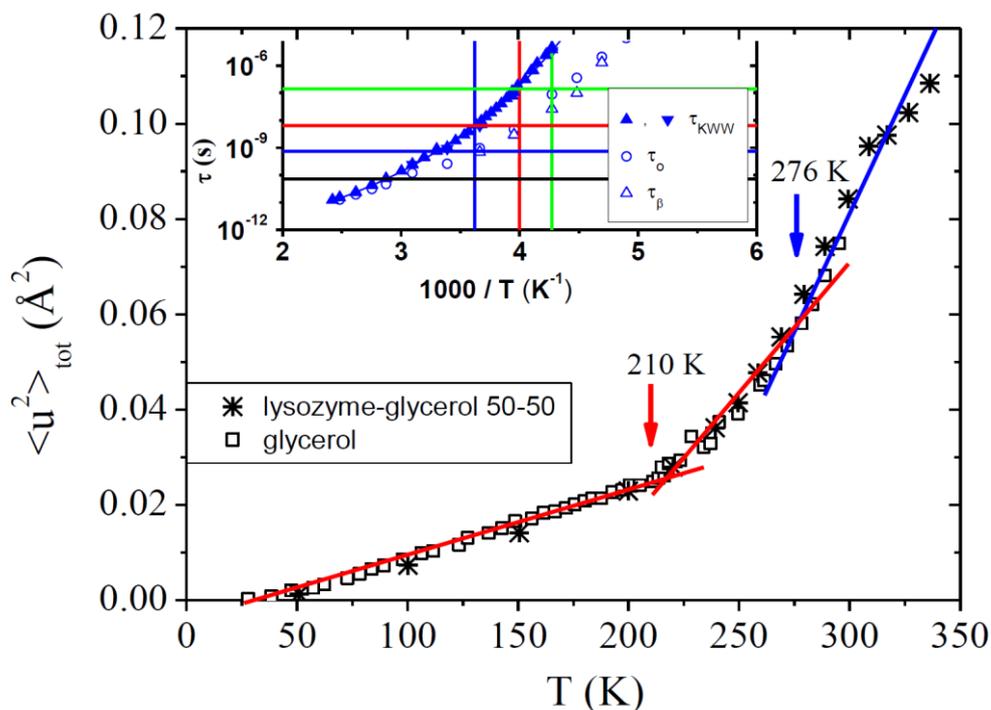

**Figure 1.** $<u^2>_{tot}$ of lysozyme solvated by only glycerol taken from Ref.[20] with additional data at higher temperatures included to compare with the mean square displacements of pure glycerol taken from Ref.[67]. Continuous lines are guides for eyes. In the inset are Kohlrausch-Williams-Watts α-relaxation time, $\tau_{KWW}$, the JG β-relaxation times obtained from a fitting procedure, $\tau_\beta$, and the primitive relaxation times, $\tau_0$, obtained from dielectric relaxation measurements from Ref.[79, 80]. The green, red, blue and black horizontals lines indicate the time-scales predicted for a relaxation process originating the dynamic transition as seen by Mössbauer spectroscopy, IN16, IN13, IN6 spectrometers, respectively. The green, red and blue vertical lines indicate the reciprocal temperatures 1000/(234 K), 1000/(250 K), and (1000/276 K) where a dynamic transition has been found in pure glycerol by means of by Mössbauer spectroscopy [82], IN16 [68], IN13 [67], respectively.

Furthermore, the similarity of $<u^2(T)>$ shows that the protein dynamics in lysozyme-glycerol (50:50) is strongly coupled to that of the solvent glycerol, in accord with the same conclusion from investigations in other solvated proteins such as hydrated maltose binding protein by Wood et al. [25]. The coupling naturally implies the possibility that the dynamic transition of lysozyme-glycerol (50:50) also can be found in bulk glycerol and at the same temperature. To look for this possibility, we need to find the common dynamic transition temperature $T_d$ where lysozyme-glycerol (50:50) undergoes the dynamic transition, and a similar 'transition' of pure glycerol. Help in answering this question can be drawn by examining the data of lysozyme-glycerol (50:50) in conjunction with those of pure glycerol obtained on IN13 spectrometer [67] reproduced separately for clarity in Fig.2, where there are shown the practically identical $<u^2(T)>$ of the partially deuterated bulk glycerol $C_3H_5(OD)_3$ and $C_3D_5(OH)_3$. Dynamic transition of lysozyme without glycerol, hydrated at the level of 0.4 g $D_2O$ g$^{-1}$, has been found by IN16 to be 220 K [24]. Since the dynamics of lysozyme solvated by glycerol is slower than hydrated lysozyme, and the fact that IN16 accesses motions longer than IN13, we can expect that $T_d$ of lysozyme-glycerol (50:50) as estimated on the latter spectrometer will be significantly higher than 220 K. In fact, a plausible location of $T_d$ of lysozyme-glycerol (50:50) is suggested by the arrow pointing at 276 K in Fig.1. The lines drawn do not carry any meaning other than used to suggest a change of $T$-dependence of $<u^2(T)>$ occurring at 276 K. Moreover, the possibility that there is change of $T$-dependence of $<u^2(T)>$ of pure glycerol occurring 270 K is suggested by the blue arrow in Fig.2. The line in the range of higher temperatures is just a guide to indicate a stronger $T$-dependence, and has no meaning because the $<u^2(T)>$ slope is continuously increasing. Summarizing, IN13 data of $<u^2(T)>$ for glycerol in Fig.2 show two changes of $T$-dependence, one around $T_g$ and another at higher temperature.

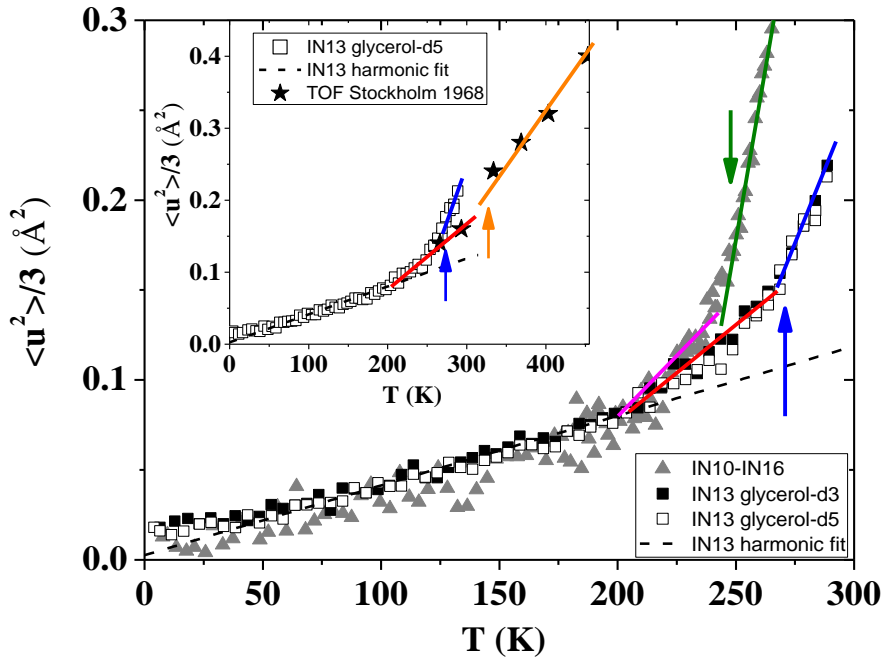

**Figure 2.** $<u^2>$ versus temperature for partially deuterated glycerol (solid (d3) and open (d5) square symbols) obtained by using IN13 [67] and for glycerol (gray triangles) by using IN10-IN16 [68]. Dashed line is the harmonic fit for the glassy behavior of IN13 data provided in ref.[67]. The solid lines are guides for eye. The green and blue arrows indicate $T_d$~250 K and 270 K, respectively. Inset plot compares IN13 data for glycerol-d5

(same symbols, lines and arrows as in main figure) with those (solid stars) obtained by means of the TOF Stockholm spectrometer [81]. Orange arrow is at $T_d$~330 K.

It is interesting to note that $<u^2(T)>$ of glycerol measured with another neutron spectrometer (IN10-IN16 [68]), sensitive to motions at longer time scales, show a similar scenario with two transitions: that at lowest temperature, occurring again around 200 K, and the other at highest temperature at $T_d$~250 K (indicated by the green arrow). On the other hand, the inset of Fig.2 shows the comparison of IN13 data of glycerol with data obtained on a shorter time scale (by means of the TOF Stockholm, $\Delta E$~200 μeV [81]). Again a transition can be noted around 200 K, but after that data continue with a linear behaviour until $T_d$~ 330 K, where a new regime begins.

The purported dynamic transition of lysozyme-glycerol (50:50) measured by IN13 at $T_d$=276 K (Fig.1) has support also from the result of Tsai et al. [19] obtained in lysozyme-glycerol (50:50). Using a spectrometer with accessible time range longer than IN13, Tsai et al. found $T_d$=265 K [84], consistent with the slightly higher $T_d$=276 K suggested in Fig.1, and very similar to $T_d$~250 K found by Niss et al. [68] that used a spectrometer with a comparable resolution (shown in Fig.2). The current estimate of 276 K for $T_d$ of lysozyme-glycerol (50:50) in Fig.1 contrasts with the much lower estimated value of $T_d$=238 K obtained in the past [20] as the intercept between the low-$T$ curve and the straight line approximating $<u^2(T)>_{tot}$ at higher temperature shown again in Fig.3. However, based on the data in an extended temperature range also from IN13 presented here in Fig. 1, we have already clearly shown the presence of the transition at about 276 K. Previously this discrepancy in the values of $T_d$ of lysozyme-glycerol (50:50) obtained by the two groups was rationalized by the use of spectrometers with different dynamic ranges and different methods of analyses of data. This rationalization is no longer needed by rediscovering that $T_d$ is 276 K from IN13 in Fig.1 as compared with 265 K from Tsai et al. [19].

Figure 3 shows the previous practice [20] of obtaining $T_d$ as the intercept between the low-$T$ solid curve and the straight line approximating $<u^2(T)>_{tot}$ at higher temperature not only for dehydrated lysozyme-glycerol (50:50) but also hydrated to different levels of $h$. All the samples were prepared with fully deuterated glycerol and heavy water so that the neutron scattering comes mainly from nonexchangeable protein hydrogen atoms. The solid curve represents the fit of the temperature dependence of $<u^2(T)>_{tot}$ by a set of quantized harmonic oscillators as in an Einstein solid [20]. The values of $T_d$ determined in this manner are 211, 207, 202 and 160 K for $h$ = 0.2, 0.35, 0.42 and 0.80 respectively. They are lower than $T_d$=220 K for lysozyme hydrated to the level of 0.4 g $D_2O$ $g^{-1}$ and without glycerol found by IN16 [24]. This cannot be true, because we expect that at least for hydration degrees less than or comparable to 0.4$h$ the presence of glycerol should make the protein dynamics slower than that of lysozyme powders hydrated at 0.4$h$, with a consistent $T_d$ higher than 220 K. This reasoning is also strengthened by the fact that IN16 can access longer relaxation times than IN13. On the other hand, the arrows, except the one labeled 238 K, indicate $T_g$ of mixtures of glycerol and water for $h$ = 0.0, 0.2, and 0.42 (from right to left) [20, 83]. Interestingly, for each hydrated lysozyme-glycerol (50:50) at level $h$ ranging from 0.2 to 0.80, $<u^2(T)>$ starts to rise above the solid line at temperature near $T_g$ of the mixture of glycerol with water having the exactly the same value of $h$. By this observation and assuming that the dynamics of hydrated lysozyme-glycerol (50:50) is strongly coupled to the solvent as in the anhydrous case shown in Fig.1 and also by other experiments [25], we can conclude that the change of *T*-dependence of $<u^2(T)>$ at $T_g$ has also been found in each of the hydrated lysozyme-glycerol (50:50). Taking this event of $<u^2(T)>$ at $T_g$ into consideration, the dynamic transition occurs at a temperature $T_d$ higher than $T_g$.

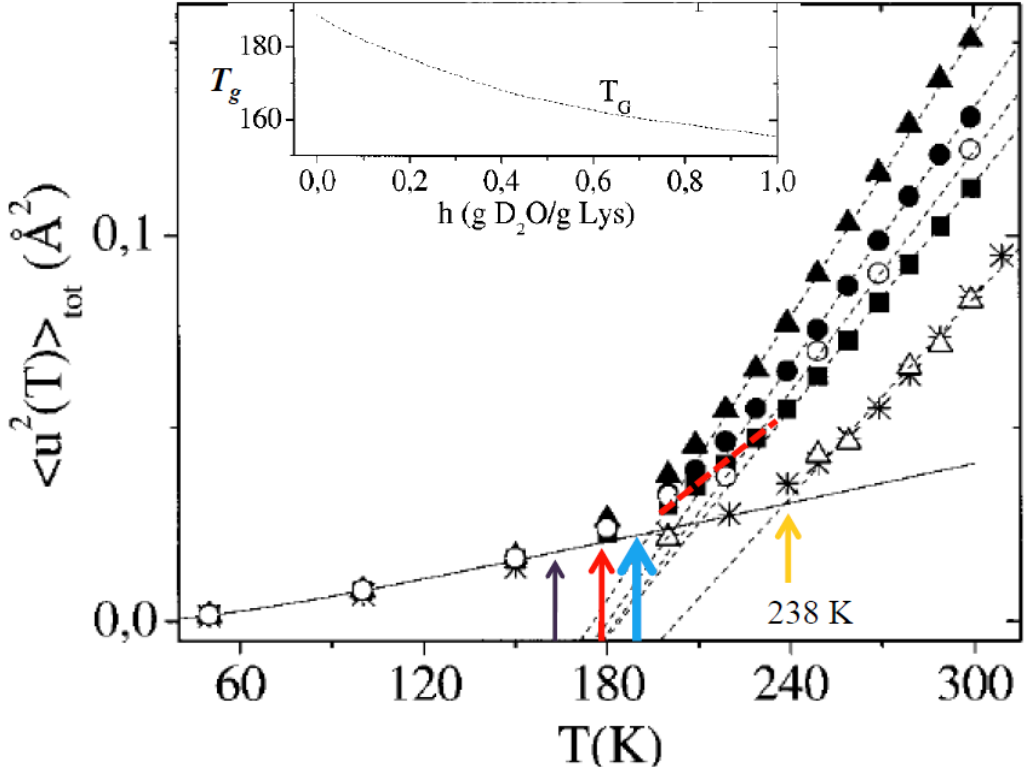

**Figure 3.** $<u^2(T)>_{tot}$ versus *T*, for all lysozyme:glycerol (50:50) hydrated at the measured water contents of: $h = 0.0$ g $D_2O$/g Lys (Δ); $h = 0.1$ g $D_2O$/g Lys (✱); $h = 0.2$ g $D_2O$/g Lys (■); $h = 0.35$ g $D_2O$/g Lys (O); $h = 0.42$ g $D_2O$/g Lys (●); $h = 0.83$ g $D_2O$/g Lys (▲). (The solid black line represent the fit of the temperature dependence of $<u^2(T)>_{tot}$ by a set of quantized harmonic oscillators as in an Einstein solid. The dashed black lines are linear fits to the high-temperature data. The intercept of the two lines for $h = 0.0$ g $D_2O$/g Lys (Δ) is at 238 K, previously considered as the dynamic transition temperature $T_d$ in Ref.[20]. The thick red and broken line through some data points of sample with $h=0.20$ is drawn to suggest $T_d$ is about 240 K, when considered together with the black dashed straight line at higher temperatures. The inset shows the $T_g$ of mixtures of glycerol and water as a function of *h*. In the main figure, the arrows from right to left indicate $T_g$ of mixtures of glycerol and water are for $h = 0.0$ (light blue), 0.2 (red), and 0.42 (purple). Readapted from ref. [20] by permission.

We have suggested that the IN 13 $<u^2(T)>$ data of pure glycerol [67] in Fig.2 show not only the change of *T*-dependence above $T_g$ but also another one at a higher temperature $T_c$ near 270 K close to the dynamic transition temperature $T_d \approx 270$ K found for the lysozyme solvated by glycerol (50:50) [20]. Does this near coincidence of $T_c$ of the solvent with $T_d$ of the solvated lysozyme also hold in the case of hydrated lysozyme:glycerol (50:50)? Answer to this question can be given for the solvent, glycerol with 15 wt% of water, which has been studied by quasielastic neutron scattering, and $<u^2(T)>$ determined at 100 ps time scale comparable to that of IN13 [44]. The data are reproduced in Fig.4. The blue arrow therein indicates the location of $T_g$ determined for glycerol with 15 wt% of water (corresponding to $h=0.18$) from the dependence of $T_g$ on water content given by Ref.[83]. The broken blue line connecting the few black data points suggests there is a change of *T*-dependence of <*u*²(*T*)> near $T_g$. The black arrow indicates the higher temperature $T_c \approx 240$ K at which <*u*²(*T*)> exercises another change in *T*-dependence. Is $T_c$ near $T_d$ of the <*u*²(*T*)> of hydrated lysozyme:glycerol (50:50) at the same *h*? The possibility that this is indeed the case is suggested by the intercept of the two lines fitting the data of the sample with *h*=0.20 at temperatures above and below ~240 K in Fig.3.

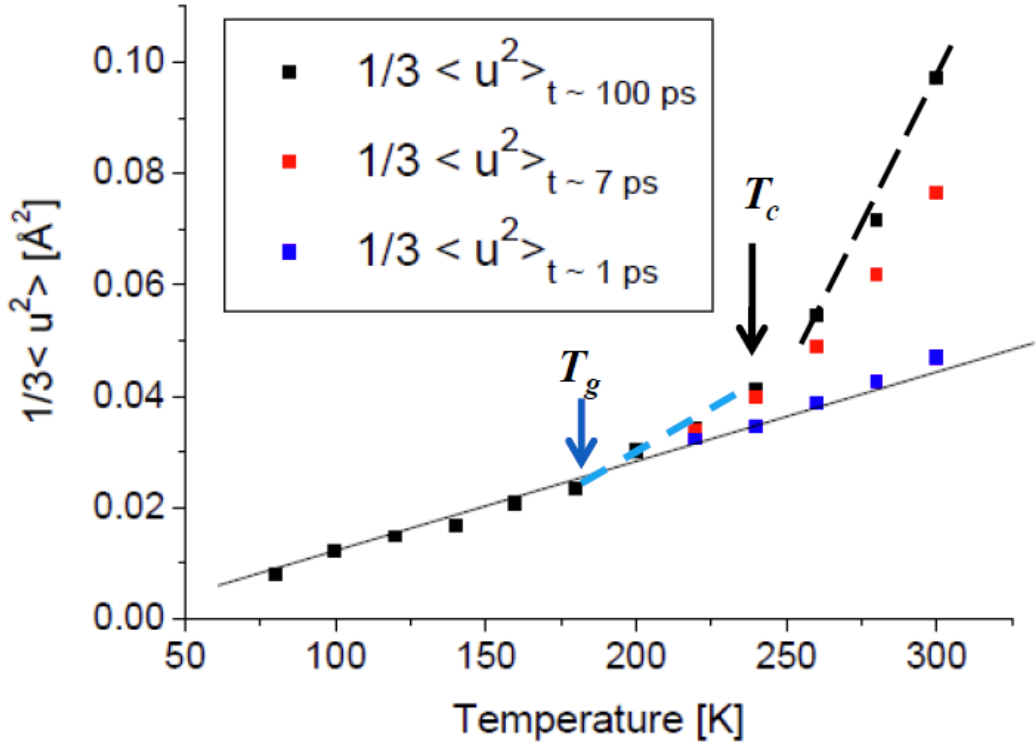

**Figure 4.** Temperature dependence of the mean square displacement in the mixture, glycerol-15 % water, as determined for different averaging times from TOF neutron scattering data by Mezei et al. [44]. The long straight line is guide to the eye. The value of $T_g$ is indicated by the blue arrow. The dashed blue and black lines are drawn to suggest change of *T*-dependence of <*u*²(*T*)> at $T_g$ and at $T_c$. The location of $T_c$ indicated by the black arrow is only suggestive because the few data points available do not allow determination with accuracy. Reproduced by permission from Ref [44].

Let us return to the inset of Fig.1 where we extract the dielectric relaxation times, $\tau_\alpha$ (labeled as $\tau_{KWW}$ in the figure) and $\tau_\beta$ of the α- and β-relaxations in the shorter time regime of pure glycerol from Ref.[79, 80]. As an equivalent estimate of β-relaxation times we can also consider the primitive relaxation times $\tau_0$ predicted by Coupling Model [53, 54], that are not affected by deconvolution fitting procedures. The three vertical lines indicate locations of 1000/*T* for *T*≈276 K (blue), 250 K (red), and 234 K (green), corresponding respectively to $T_d$ of lysozyme:glycerol (50:50) in Fig.1, $T_c$ of pure glycerol from IN10-IN16 spectrometer in Fig.2, and the value for crossover temperature obtained from Mössbauer spectroscopy [82], that, by the way, is close to $T_d$≈238 K obtained by the intercept of the black dashed and continuous lines in Fig.3 for lysozyme:glycerol (50:50) in ref.[20].By interpolation, we can estimate the values of $\tau_\beta$ or $\tau_0$ at $T \approx 276$ K, 250 K, and 234 K to be around 800 ps, 6 ns and $5 \times 10^{-8}$ s. According to recent studies [34], the change of T-dependence of $<u^2(T)>$ at $T_d$ is originated when the time scale of a relaxation process becomes shorter than the longest time scale detectable by the neutron spectrometer (for neutron scattering $\tau_{res}$ is defined in caption of Table 1, for Mössbauer spectroscopy is 140 ns). More precisely, if the relaxation process has a Lorentzian shape, the onset of the transition at $T_d$ occurs when $\tau_{res} \sim 0.2\tau_{max}$ [34], where $\tau_{max}$ is the most probable time of the relaxation process (that could be α-, β-, or their merging). The horizontal black, blue, red, and green lines in Inset of Fig.1 indicate when the above predicted relation should occurs for different experimental apparatuses as IN6, IN13, IN16 spectrometers and by means of Mössbauer spectroscopy, respectively. By inspection, $\tau_\beta$ or $\tau_0$ falls very close to the expected predictions. Hence these can explain the dynamic transition at $T_d$ =276 K for solvated lysozyme and at $T_c$ =270 K for pure glycerol, both by IN13 with time window of the order of hundreds of ps, by the β-relaxation entering the window of the spectrometer on raising temperature to cross $T_d$ and $T_c$. In the same way, it could explain the $T_c$ =250 K seen by Niss et al. [68] for glycerol using IN10 (Fig.2) and the very high value for transition (>330 K) seen by Larrson using a spectrometer with a very limited resolution (see Inset Fig.2). This explanation of the dynamic transition follows the same way the phenomenon was explained in hydrated proteins, namely the water-specific or the Johari-Goldstein β-relaxation of the hydration water given before in Ref.[41, 85, 42, 43, 60]. There, it has been shown that on increasing temperature, the Johari-Goldstein β-relaxation of the hydration water starts entering the time window of spectrometer, near $T_d$, leading to stronger increase of $<u^2(T)>$ with temperature and the dynamic transition. On the other hand, $\tau_\beta$ at $T \approx 238$ K is $\sim 5 \times 10^{-8}$ s, which is more than a factor 70 longer than the time window of IN13 and hence β-relaxation of glycerol cannot cause the dynamic transition in anhydrous lysozyme:glycerol (50:50) at 238 K [20]. This is another reason for not accepting 238 K as the temperature of the occurrence of the dynamic transition.

As for the universal changes of T-dependence of the fast process at $T_g$ measured in terms of $<u^2(T)>$ by neutron scattering [62], susceptibility $\chi''(\omega)$ by light scattering [50, 86], and dielectric loss $\varepsilon''(\omega)$ by dielectric spectroscopy [55], the first question to ask is the identity of the fast process, and the second question to follow is why it is sensitive to glass-liquid transition at $T_g$. Answers to these questions have been given before in Refs.[47, 62, 55], and will be discussed further in section 2.5.

*2.2 Hydrated Lysozyme*

Sometimes the task of finding the change of T-dependence of $<u^2(T)>$ at $T_g$ in hydrated myoglobin and lysozyme is hampered by the contribution from the rotation of the methyl group present. The relaxation time of methyl group rotation has an Arrhenius temperature dependence with an average activation energy of 10.5 kJ/mol [34, 63, 66, 69, 87, 88]. Some of these neutron scattering studies have found the onset of anharmonicity in samples at all hydration level in increasing temperature starting at $T \sim 100$-150 K and continuing to higher temperatures. Since at higher levels of hydration of the solvent, the $T_g$ of the pure solvent as well as the corresponding solvated proteins can reach temperatures not much above 150 K, [7, 89, 9, 10, 11, 12, 13, 14] existence of the change of T-dependence of $<u^2(T)>$ at $T_g$ in these samples can no longer be ascertained. An example of this situation is presented in Fig.5 for lysozyme hydrated at 0.4*h*. taken from the published data of the total mean square displacement in Ref.[69]. In the back-scattering time window of IN 13, the methyl group rotation contribution is significant, which makes it difficult to discriminate this process from the feature we are looking for near $T_g$. Certainly part of the increase in *T*-dependence of $<u^2(T)>_{tot}$ above 100 K is due to methyl group rotation. By using neutron scattering with energy resolution of 1 μeV, Roh et al. have demonstrated the onset of anharmonicity due to methyl group rotation in dry and hydrated lysozyme is at 100 K independent of hydration level [66]. Although the methyl group rotation continues to contribute to $<u^2(T)>_{tot}$ at higher temperatures past the onset at 100 K, its contribution is expected to be smooth, and cannot account for the presence of the break in the temperature dependence of $<u^2(T)>_{tot}$ near 173 K (see Fig.5). Therefore, we can be certain that this break, which has been determined also by Brillouin scattering [14], is due to crossing the glass transition temperature of lysozyme at 0.4*h*. Another increase in *T*-dependence of $<u^2(T)>_{tot}$ above ~230 K can be identified with the dynamic transition with $T_d$~230 K measured by IN13. Roh et al. also found this dynamic transition in the range 200-220 K with spectrometer that has time resolution of about 1 ns, ten times better than that of IN13.

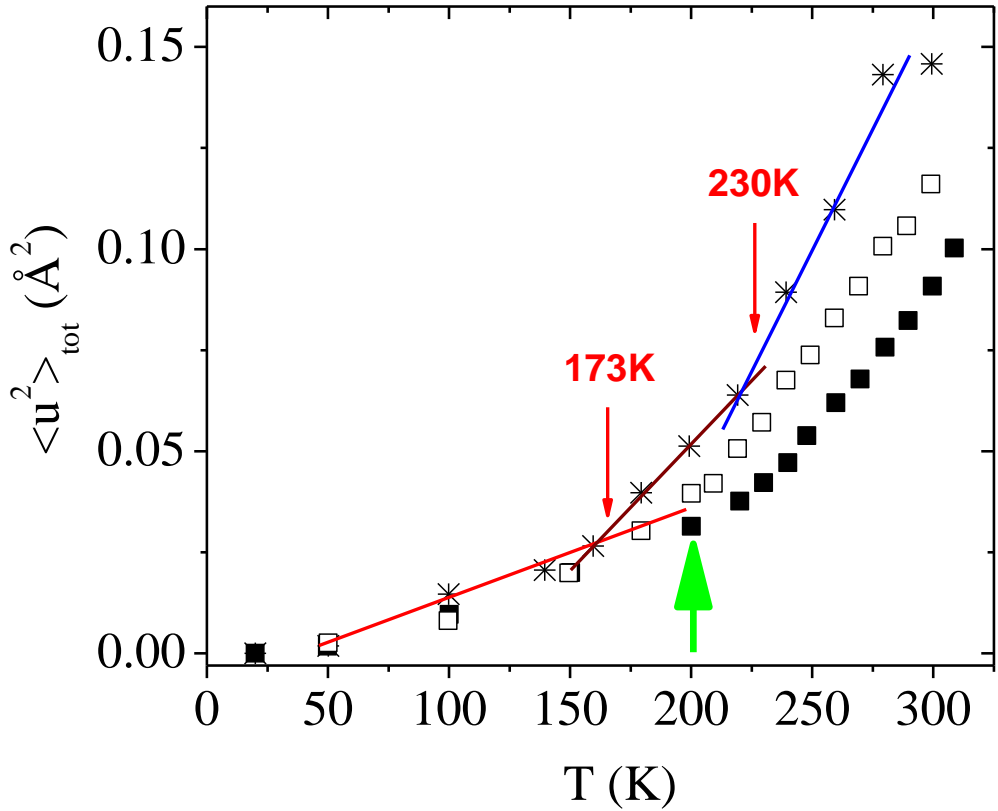

**Figure 5.** Total mean square displacements versus *T* for lysozyme at 0.4*h* (star), lysozyme in glycerol at 0.4*w* (empty square) and lysozyme in glucose 0.4*w* (full square) measured by IN 13. Data are from Ref.[69]. The red arrow indicates 173 K, which can be taken within uncertainty as the glass transition temperature of lysozyme at 0.4*h*. The lines drawn merely are used to suggest change of *T*-dependence of $<u^2(T)>_{tot}$ near 173 K, and near 230

K. The green arrow indicates $T_g\sim200\pm5$ K of lysozyme in glucose 0.4$h$ as well as the pure solvent glucose 0.4w itself.

*2.3 Lysozyme solvated by glucose-water and hydrated glucose*

The mean-square displacements of the hydrogen atoms of lysozyme solvated by deuterated glucose-water mixtures at various water contents $h$ (=$g$ D$_2$O/$g$ glucose), ranging from the dry sample with 0$w$, hydrated sample with 0.15$h$, 0.40$h$, 0.60$h$, and 0.70$h$, had been measured by elastic neutron scattering using IN13, and partially reproduced here in Fig.6 [69, 70]. The lysozyme and deuterated glucose have equal weight in the samples. The dry sample, 0$h$, as well as the hydrated samples shows an onset of anharmonicity at ~100 K and continues to higher temperatures, which is attributed to the activation of methyl group reorientations intrinsic to the protein. The contributions to $<u^2(T)>_{tot}$ from the methyl group rotation plague the task of looking for the change of $T$-dependence at $T_g$ in all the hydrated samples except the case at 0.15$h$. This is because the $T_g$ of the sample with 0.40$h$ is ~200 K and lower for 0.6$h$ and 0.7$h$, and methyl contribution dominates $<u^2(T)>_{tot}$ at these low temperatures. On the other hand, one clear case that is not plagued by the methyl group rotation is the sample with 0.15$h$ which has a much higher $T_g\approx242$ K [9, 89], and the measured $<u^2(T)>_{tot}$ rises above that of the dry sample at temperatures above ~250 K (see Fig.6), and hence the data at higher temperatures cannot come from the methyl group. Part of this rise was considered before as the signature of the dynamic transition at $T_d\approx260$ K [70]. However, dielectric relaxation measurements of glucose with water content close to 0.15$w$ have been made recently [90]. The fastest relaxation observed is from the water component in the mixture, but at 260 K its relaxation time $\tau_\beta$ is several orders of magnitude longer than 1 ns or 100 ps at 260 K. Thus no relaxation process of the solvent can enter the time window of IN 13 at 260 K. An estimate of the temperature at which $\tau_\beta$ reaches 1 ns is 310 K, and hence $T_d$ must be 310 K or higher, which is outside the temperature range of measurements shown in Fig.6. After eliminating both the dynamic transition and the methyl rotation as the cause of the change of $T$-dependence of $<u^2(T)>_{tot}$ near 240-250 K, the natural choice remaining is the general property of the change of $T$-dependence of $<u^2(T)>_{tot}$ near $T_g$ of the sample with 0.15w.

The dynamic transition temperature $T_d$ of the lysozyme in glucose at 0.4$h$ was determined from the data of $<u^2(T)>_{tot}$ in Fig.6 to be 220 K [70]. This value of $T_d$ is lower than that of hydrated lysozyme without glucose by IN13 (230-240 K, see Fig.5) This is unreasonable considering that the presence of glucose in the 0.4$h$ sample will slow down the dynamic and hence $T_d$ of the lysozyme in glucose at 0.4$h$ has to be significantly higher than 230-240 K of hydrated lysozyme with 0.4$h$. Dielectric data [90] of glucose with water content close to 0.4$h$ also show $\tau_\beta$ is about $10^{-6}$ s at 220 K, and hence the dynamic transition of the 0.4h sample is not at 220 K but has to be higher than 280 K, a figure obtained by extrapolating dielectric data to shorter time. By the same reasoning as given before for the 0.15$w$ sample, it seems that the change of $T$-dependence of $<u^2(T)>_{tot}$ near $T_g\approx200\pm5$ K of the sample with 0.4$h$ from dielectric relaxation [90] also exists in the data (closed triangles) of Fig.6. The reader can find in Fig.5 an unobstructed view of the same data and the change in $T$-dependence of $<u^2(T)>_{tot}$ near $T_g\approx200\pm5$ K.

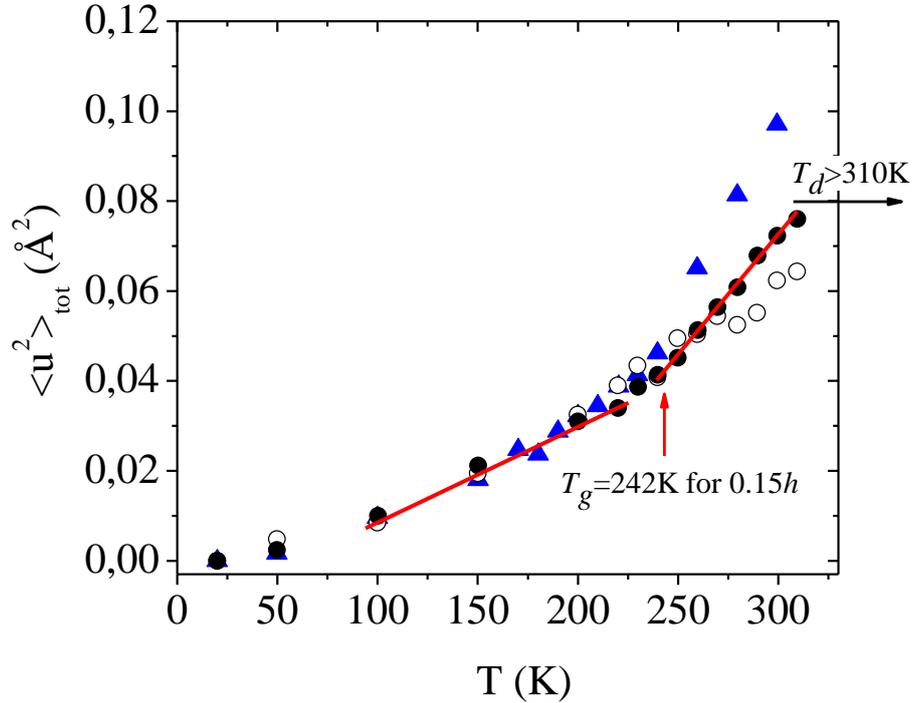

**Figure 6.** Total mean square displacement of lysozyme in deuterated glucose at $0h$ (O), $0.15h$ (●), $0.4h$ (▲), $0.6h$ (✱), and $0.7h$ (Δ). The arrow indicates $T_g$= 242 K, the glass transition temperature of the sample with $0.15h$. The lines serve no purpose except to bring out the change of $T$-dependence of $<u^2(T)>_{tot}$ at some temperature near $T_g$. Data from Ref.[70]

Measurements of $<u^2(T)>_{tot}$ of lysozyme in deuterated glucose at $0h$ shown in Fig.6 had been extended to higher temperatures also by IN 13. These previously unpublished data are presented in Fig.7. Over the expanded temperature range, $<u^2(T)>_{tot}$ exhibits two breaks in its $T$-dependence. The one at lower temperature nearly coincides with $T_g$=307.6 K of pure glucose [91]. This is the manifestation of the change of $<u^2(T)>_{tot}$ at $T_g$ of lysozyme solvated by deuterated glucose can be identified. Here we have assumed that the $T_g$ of lysozyme in deuterated glucose is nearly the same as $T_g$ of deuterated glucose, which is supported by the findings of dielectric experiments [90]. The next break in $T$-dependence of $<u^2(T)>_{tot}$ can be identified with the dynamic transition which occurs at $T_d$ near or above 370 K. Dielectric relaxation data of Kaminski et al. [91] obtained up to near 1 GHz show the secondary relaxation time of glucose is about 1 ns at 370 K. The relaxation time of this secondary relaxation is one decade longer than the IN13 time window of 100 ps. However, considering the fact that the frequency dispersion of the secondary relaxation is very broad, it is plausible that the dynamic transition is caused by the secondary relaxation entering the time window of IN13 starting from 370 K. The advantage of acquiring data of $<u^2(T)>_{tot}$ at higher temperature for positive identification of the dynamic transition is clearly demonstrated by the example of lysozyme solvated by glucose. Had this not been done, the dynamic transition at about 370 K would be missed or mistakenly identified with the break at the lower temperature near 307 K which turns out to be the crossover at $T_g$. In the inset of Fig.7, the higher temperature data of $<u^2(T)>_{tot}$ of lysozyme solvated by deuterated glycerol help to positively identify the dynamics transition and accurately determine the value of $T_d$ at 270 K. Although not many data points were obtained at lower temperatures, the data suggest an onset near 100 K which can be due to the methyl group rotation as found by Roh et al.

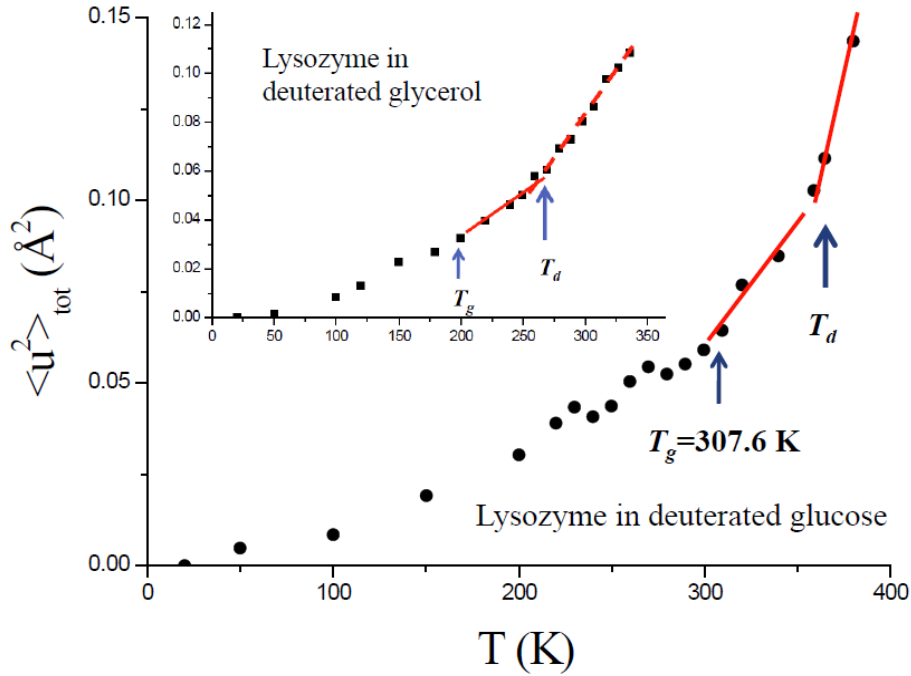

**Figure 7.** Total mean square displacements of lysozyme in deuterated glucose at $0h$ (●) measured up to high temperatures help to clearly identify the crossover at $T_g$ and the dynamic transition at $T_d$. The inset shows the same for lysozyme solvated by deuerated glycerol.

Di Bari et al. [71, 72] reported IN13 elastic neutron scattering data on the monosaccharide, glucose, and its polymeric forms, amylose and amylopectin, over the hydration range $h$ from the dry state to about 0.6 ($h$=g water/g dry saccharide). Their data of glucose with $h$=0, 0.25, and 0.50 are reproduced in Fig.8. The locations of $T_g$ for $h$=0.25 and 0.50 are marked by arrows. Again, from dielectric relaxation measurements [90, 92, 93, 94, 95] of aqueous mixtures of glucose with $h$ close to 0.25 extrapolated to higher temperatures, it is estimated that temperature higher than 300 K is needed to for $\tau_\beta$ to reach 1 ns, which is ten times longer than the time window of IN 13. Hence, the dynamic transition in the sample with $h$=0.25 has not been seen in Fig.8, and the data therein actually reflect the change of $T$-dependence of $<u^2(T)>$ near $T_g$=220 K. This interpretation of the data draws a distinction from the dynamic transition of Di Bari et al.[ 71]. In their interpretation, the temperature of activation of the hydration dependent anharmonic behavior, $T_a$, is basically an estimate of the dynamic transition temperature $T_d$. The value of $T_a$ is ≈ 178 K for $h$=0.25, remarkably much lower than $T_d \approx$230-240 K by IN13 of hydrated lysozyme or myoglobin at $h$=0.40 [7, 14] and higher at $h$=0.25, without the presence of glucose. This unreasonably low value of $T_a$ invalidates the interpretation of dynamic transition in the sample with $h$=0.25, and also in the sample with $h$=0.50 having $T_a \approx$150 K.

At values of $h$ comparable to those of the hydrated glucose discussed in the above, the $T_d$ of the hydrated polymeric amylose and amylopectin should be even higher, and lies outside the temperature range where $<u^2(T)>$ data were taken, whose upper limit was nearly 340 K. Hence, the low values of $T_a$ in the range from 210 K to 150 K of amylase and amylopectin with $h$ ranging from 0.15 to 0.60 cannot be related to dynamic transition.

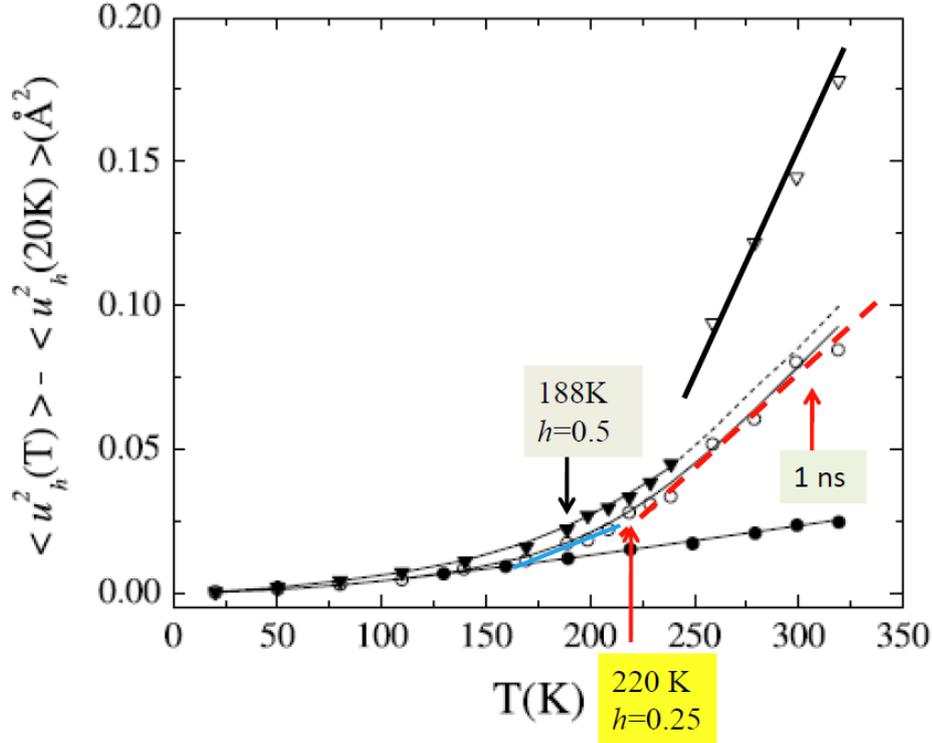

**Figure 8.** Total mean square fluctuation differences reproduced from Ref.[71] by permission. The continuous lines are fits to a harmonic Debye model for the dry sample, and to the double-well potential model for the hydrated ones. The data refer to glucose samples at hydrations $h = 0$ (●), 0.25 (O), 0.50 (▼). The data above 250 K for the highest hydration reported (open triangles) are obtained from the initial slope of the elastic temperature scans.

*2.4 Dynamic transition observed at $T_d$ and/or $T_g$ in hydrated disaccharides*

Elastic neutron scattering at IN13 has been used to study the dynamic properties of the aqueous mixtures of disaccharides, trehalose, maltose and sucrose, and their mixtures in water [73, 74, 75]. The hydrated disaccharide samples have two different molar contents of water, designated by the authors as disaccharide+6H$_2$O and disaccharide+19H$_2$O, corresponding to $0.32h$ and $1h$ respectively. The $<u^2(T)>$ data of all three disaccharides are similar in all respects including temperature dependence, and the representative one for trehalose+6H$_2$O and trehalose+19H$_2$O are reproduced in Fig.9.
In Refs.[73, 74] the authors gave the values, 238, 235 and 233 K, for $T_g$ of trehalose+19H$_2$O, maltose+19H$_2$O, and sucrose+19H$_2$O respectively, and identified the onset of the rapid rise of $<u^2(T)>$ above the harmonic response at lower temperatures to occur at $T_g$ (see Fig.9 for example) rather than at a higher temperature $T_d$, which is usually the case for dynamic transition in solvated proteins or the solvents, such as in all the examples discussed before in the present paper. One of us (KLN) [53] cited Refs.[73, 74] for evidence of change of $T$-dependence of $<u^2(T)>$ at $T_g$, considering 238 K as the true $T_g$ of trehalose+19H$_2$O, and similar values for the other disaccharide+19H$_2$O. Actually, after checking the values of $T_g$ of the disaccharide+19H$_2$O in the publications by Green and Angell [9] and by Bellavia et al. [12], it becomes clear that these values quoted in Refs. [73, 74, 75] are far above the actual values of $T_g$, as obtained from calorimetric measurements. Reading off from Figs.5a-5c of the paper by Bellavia et al. [12], the correct $T_g$ are about 170-172 K for all three disaccharide+19H$_2$O. So, how to explain this apparent conundrum, i.e. the appearance of the onset of rise of trehalose+19H$_2$O, maltose+19H$_2$O, and sucrose+19H$_2$O at 238, 235 and 233 K respectively, occurring about 65 degrees above $T_g$? One could invoke the presence of two distinctive $T_g$ in this samples, one for the hydration water (lower) and the other for the disaccharides, but this is at odds with the accurate calorimetric measurements [12] and not common for hydrogen bonded mixtures, unless to invoke a phase separation. Actually, the simplest and more reasonable identification for these onset temperatures is that they are likely the dynamics transition temperatures of the disaccharides with 19H$_2$O because from dielectric relaxation data [90] of trehalose+6H$_2$O the $\tau_\beta$ has already reached ~$10^{-7}$ s at 238, and increase to 19H$_2$O will easily take it to the time scale of 100 ps to 1 ns for detection by IN13. Recently, Gabel and Bellissent-Funel [26] have found the dynamic transition of hydrated C-phycocyanin in trehalose+26H$_2$O (equivalent to 1.36$h$) by incoherent elastic neutron scattering using IN 13 and IN 16. In the temperature range from 20 to 230 K, the measured $<u^2(T)>$ from both spectrometers were identical and increased with temperature nearly linearly. Then, at 240-250 K the rapid increase of $<u^2(T)>$ indicates the dynamical transition. On the other hand, the dynamical transition observed on IN16 was shifted to a slightly lower temperature than the one observed on IN13, due to the longer time window of the former.

Notwithstanding the temperature values given in [73, 74, 75] for the disaccharide+19H$_2$O are to be ascribed to $T_d$ and not to $T_g$, the values for the disaccharide+6H$_2$O (corresponding to 25 wt%) are, on the contrary, almost exactly the same as the values of $T_g$ accurately determined from calorimetry data [9, 12]. Reading off from the chart of Bellavia et al., $T_g$ of trehalose+6H$_2$O is ≈ 230 K, which is nearly the same temperature at which the measured $<u^2(T)>$ exhibit a rapid rise from the harmonic fit shown in the inset of Fig.9. Therefore, the change of $T$-dependence of $<u^2(T)>$ at $T_g$ is indeed observed in trehalose+6H$_2$O. By the way, in mixtures of trehalose with 25 wt% of water, close to composition of trehalose+6H$_2$O, the $\tau_\beta$ from dielectric data [90] is about $10^{-6}$ s, which is more than 3 orders of magnitude longer than the time window of IN13 and hence cannot give rise to dynamic transition. An extrapolation of the dielectric $\tau_\beta$ to higher temperature suggests that it will reach the time window of IN13 at temperatures higher than about 300 K, outside the temperature range of measurements shown in the inset of Fig.9. In this case the higher onset temperature $T_d$ cannot be observed as it is outside the experimentally investigated temperature range. In the case of trehalose+19H$_2$O (main figure in Fig.9) the sharp rise is to be ascribed to $T_d$ and probably overwhelms the milder change at $T_g$, occurring at lower temperatures.

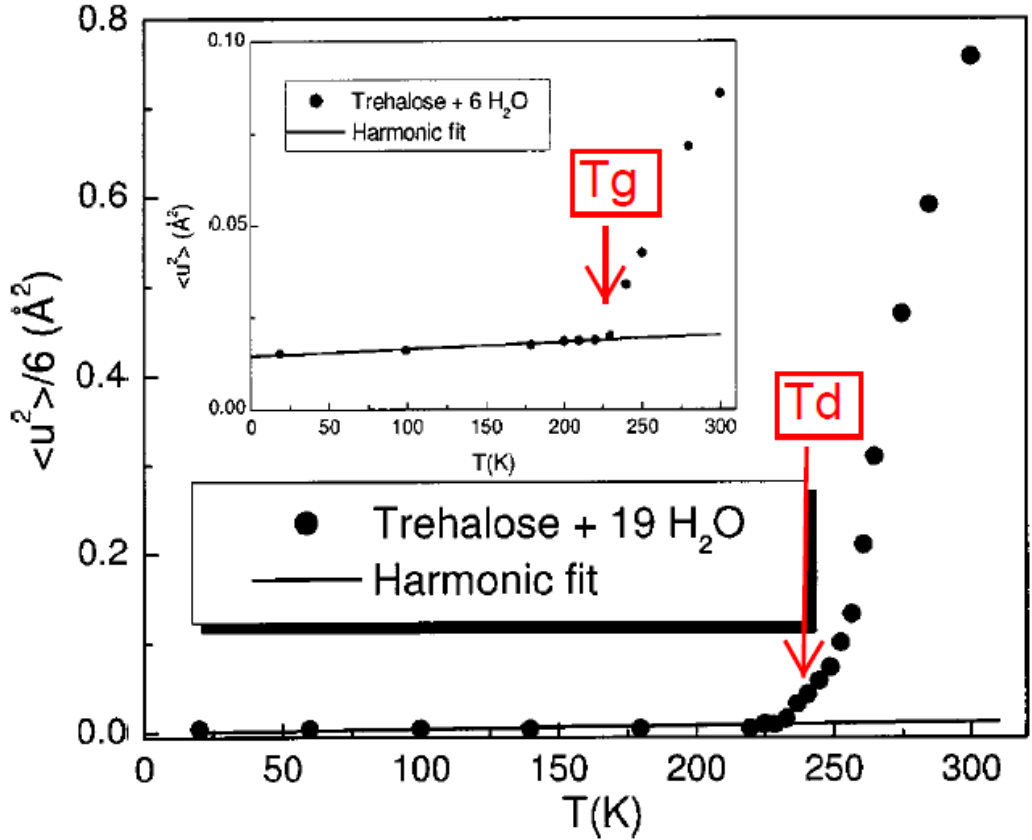

**Figure 9.** Figure reproduced from Ref. [73] show the dependence of $<u^2(T)>$ on temperature for trehalose+19H$_2$O in the main figure, and trehalose+6H$_2$O in the inset. The solid lines are the harmonic fits. For trehalose+19H$_2$O, the change of T-dependence at $T_d \approx 238$ K ($>T_g \approx 170$ K) is the dynamic transition. On the other hand, for trehalose+6H$_2$O, the change of T-dependence is at $T_g \approx 230$ K. Reproduced and adapted from Refs.[73] and [75] by permission.

*2.5 Origin of the change of $<u^2(T)>$ at $T_g$ in neutron scattering experiments*

The main theme of the present paper is on the presentation of evidence for the change of $T$-dependence of $<u^2(T)>$ at $T_g$, preceding the dynamic transition of $<u^2(T)>$ at a higher temperature $T_d$. Nevertheless, it is worthwhile to briefly mention here the origin of the change of $T$-dependence of $<u^2(T)>$ at $T_g$. Despite this is a universal phenomenon found in many different classes of glass-formers [47, 48, 49, 51, 54], so far there is only one attempt to elucidate its origin [52, 53, 55, 62] as far as we know. The $<u^2>$ obtained by neutron scattering at not so low temperatures has contributions from the dissipation of excursion of molecules while caged by the anharmonic intermolecular potential at times before the cages are dissolved by the onset of β-relaxation [54] or by the merged αβ-relaxation. In the present case of hydrated protein, the β-relaxation is the JG β- or water-specific relaxation of water in the hydration layer [41, 42, 43, 60], while in solvated protein without water is the JG β-relaxation of the solvent itself like glycerol in Fig.1. Unlike genuine relaxation process, the dissipation of caged molecules has no characteristic relaxation time, and hence it appears as a "nearly constant loss (NCL)" well approximated in the susceptibility spectrum $\chi''(\omega)$ or the dielectric loss function $\varepsilon''(\omega)$ by a power law, $A\omega^{-c}$, where $c$ is small and positive. Correspondingly, the mean square displacement, $\langle u^2(t)\rangle$, as a function of time has the power law dependence, $Bt^c$, with $c\ll 1$.

This feature is commonly seen in conventional glass-formers, and it has been seen also in maltose binding protein (MBP) hydrated to about one hydration layer per MBP molecule by molecular dynamics simulation, as it is shown in the main part of Fig.10 reproduced from Fig.2b of Wood et al. [25]. Presented are the time evolutions of mean-square displacements of water O atoms over a range of temperature from 150 to 300 K. The authors reported that 240 K is the temperature $T_d$ of the dynamic transition in the simulations for times up to 100 ps. These times are comparable with the time window of the IN13 spectrometer, using which $T_d$=240 K is often observed on fully hydrated proteins. The temperature dependence of the mean square displacements of the O atoms of the water molecules at 100 ps is presented in the inset of Fig.10. In addition to the dynamic transition at $T_d$=240 K, the mean square displacements show a change of $T$-dependence at about 200 K in the simulation, which corresponds to the 4$^{th}$ curve from the bottom in the main figure marked by an arrow. We take this finding by simulations as another clear evidence of the presence of the change of $T$-dependence of $\langle u^2(T)\rangle$ at $T_g$.

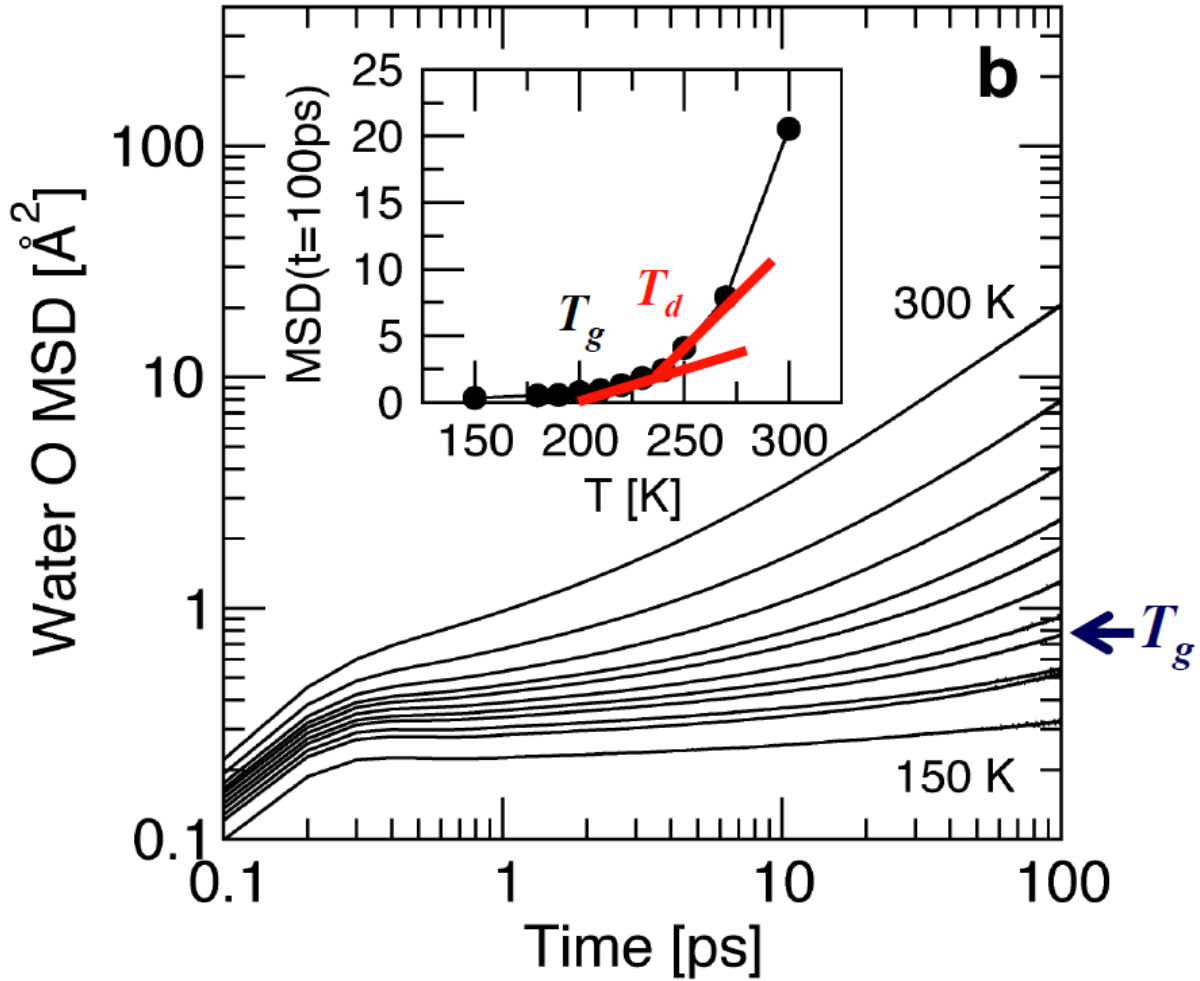

**Figure 10.** Time evolution of mean-square displacements of water O atoms in hydrated MBP over a range of temperature from 150 to 300 K (bottom to top: 150, 180, 190, 200, 210, 220, 230, 240, 250, 270, 300 K; the curve at 240 K, the temperature of the dynamical transition in the simulations, is indicated by a heavy line).

(Inset) Temperature dependence of the water MSDs at 100 ps. In addition to the dynamic transition at $T_d$=240 K, the MSDs show a change of $T$-dependence at about 200 K in the simulation, which corresponds to the 4$^{th}$ curve from the bottom in the main figure and is marked by the arrow . Reproduced from Ref.[25] by permission

The identification of change of $T$-dependence of $<u^2(T)>$ of the hydration water at $T_g$ in purple membrane (PM) can resolve a conundrum resulting from the recent study by neutron scattering and diffraction experiments on hydrated stacks of purple membranes by Wood et al. [76, 77]. In this study, they examined the dynamical coupling between the PM and the hydration water by a combination of elastic incoherent neutron scattering, specific deuteration, and molecular dynamics simulations. The dynamics of hydration water were isolated by measurements on completely deuterated PM and hydrated in $H_2O$, while the PM dynamics were obtained from the study of natural abundance PM in $D_2O$. The temperature-dependence of $<u^2(T)>$ shows changes at 120 K and 260 K for the PM, and at 200 K and 260 K for the hydration water. An outstanding difference in the dynamics of PM and hydration water is brought out by the presence of change of $T$-dependence of $<u^2(T)>$ of hydration water at 200 K, in contrast to the absence in the $T$-dependence of $<u^2(T)>$ of PM at the same or nearly the same temperature. This observation in membrane differs from soluble proteins where the <u>dynamic transition</u> can be observed by the change of $T$-dependence of $<u^2(T)>$ of either the hydration water or the protein at the same temperature. The findings in hydrated PM have casted doubt on the current view that the dynamics of protein are coupled to the hydration water and the two exhibit dynamical changes at the same temperature [25, 96]. The words, dynamic transition, in a few lines above are underlined to underscore the fact that the dynamical change of soluble proteins intended by Wood et al. is apparently the change of $T$-dependence of $<u^2(T)>$ at $T_d$, and not at $T_g$. If the change of $T$-dependence of $<u^2(T)>$ of water in hydrated PM at 200 K is indeed at $T_g$ which we profess to be general, then there is no contradiction to the current view that that macromolecular motions respond to dynamical changes in the hydration water as far as the <u>dynamic transition</u> at $T_d$ is concerned and if $T_d$=260 K. This is because both water and membrane show change of $T$-dependence of $<u^2(T)>$ at 260 K. Corroborating evidence of this can be found from earlier elastic incoherent neutron scattering study of dry hydrogenated PM and $D_2O$-hydrated hydrogenated PM using IN10 by Réat et al. [78] with energy resolution $\Delta E$=1 μeV and the $<u^2(T)>$ values corresponding to motions occurring in a time shorter than 2 ns. The $D_2O$-hydrated hydrogenated PM shows the change in $T$-dependence of $<u^2(T)>$ or the dynamic transition at $T_d$ near 270 K, labelled as 'solvent effect' by Réat et al. On the other hand, no such dynamic transition was found in dry hydrogenated PM. The fully deuterated PM and hydrated by $D_2O$ studied by Réat et al. also show the dynamic transition at $T_d$ in the range of 260-270 K. All the present and previous studies of PM support that the observed change in $T$-dependence of $<u^2(T)>$ near 260 K is indeed the dynamic transition of the coupled protein-hydration water and the two exhibit dynamical changes at the same temperature $T_d$.

Actually there is experimental evidence that $T_g$ of the hydrated PM studied by Wood et al. is about 200 K. Berntsen et al. [97] had studied PM hydrated to $h$=0.4 and 0.2g $H_2O$/g of PM by dielectric spectroscopy and differential scanning calorimetry between 120 and 300 K. They found by calorimetry a pronounced endothermic process at 190-200 K, which can be identified as $T_g$ of the hydrated PM. By dielectric spectroscopy they found the JG β-relaxation of the hydration water with relaxation time $\tau_\beta$ showing Arrhenius $T$-dependence at low temperatures with activation energy of about 54 kJ/mol typical of this kind of relaxation, and changing to a stronger *T*-dependence at high temperatures after crossing 190–200 K. This behavior is another indication that $T_g$ is within the range of 190-200 K because change of *T*-dependence of $\tau_\beta$ at $T_g$ is a universal feature found in all glassformers including aqueous mixtures and hydrated proteins [43, 60, 58]. Since the sample studied by Wood et al. is hydrated to the level of approximately *h*=0.3 g of water per g of PM, in between 0.4 and 0.2g $H_2O$/g of PM in the samples of Berntsen et al., we can conclude that its $T_g$ is within the range of 190-200 K. Additional support of $T_g \approx 200$ K may be drawn from the observed onset of translational mobility in water beyond the first hydration layer indirectly observed at 200 K by monitoring the lamellar spacing of PM stacks as a function of temperature. The molecular dynamics simulation of hydration water of PM has $<u^2(T)>$ at 30 ps changing *T*-dependence near 200 K.

The dielectric relaxation data of $\tau_\beta$ from Berntsen et al. enable us to verify that the transition seen at 260 K in the $<u^2(T)>$ of either the hydration water or the membrane protein is indeed the dynamic transition caused by the JG β-relaxation entering the spectral range of the IN 16 with time window shorten than about 1 ns. On extrapolating the dielectric $\tau_\beta$ obtained for *h*=0.2 and 0.4 above 200 K by an Arrhenius law, we estimate at 260 K that $\tau_\beta$ is a factor of about 4 or 5 shorter than 1 ns. The actual *T*-dependence of $\tau_\beta$ is likely weaker than the Arrhenius dependence assumed because the prefactor of the latter is about $10^{-19}$ s and is unphysical. Therefore, a good correspondence in order of magnitude between $\tau_\beta$ at *T*=260 K and the IN 16 time resolution of about 1 ns is possible.

The fact that membrane motion does not show the change of *T*-dependence of $<u^2(T)>$ at 200 K is clear. This suggests that the membrane dynamics is not as sensitive to glass transition as hydration water, probably because presence of lipids may have a stronger impact on the membrane protein dynamics than hydration water, when the glass transition temperature is crossed. This behavior is at variance with what we have seen above (see for instance Fig. 3 and Fig. 7) in some soluble proteins, where the change of *T*-dependence of the protein $<u^2>$ at $T_g$ has been seen. However, for the dynamic transition, the dynamics of the membrane protein is still coupled to or controlled by that of hydration water and it occurs at $T_d \approx 260$ K. Thus, by identifying specifically that the 200 K transition is at $T_g$, the challenge by the neutron scattering data of hydrated PM on the current view of the dynamic transition is removed. Returning to the experimental fact that the power laws of $\chi''(\omega) = A\omega^{-c}$ or $<u^2(t)> = Bt^c$, with *c*<<1 terminates at times of the order of $\tau_\beta$, it was argued that the intensity factor *A* or *B* approximately varies with temperature like $1/\log(\tau_\beta(T))$ [62]. Experimentally, it is found in conventional glass-formers [98, 99] as well as in aqueous mixtures [61, 58, 39, 100] and hydrated proteins [85, 14], that the relevant $\tau_\beta(T)$ changes from the Arrhenius *T*-dependence below $T_g$ to a stronger *T*-dependence above $T_g$. Therefore, this change in *T*-dependence of $\tau_\beta(T)$ on crossing $T_g$ leads to the corresponding change in *T*-dependence of $<u^2(T)>$ at $T_g$. Conceptually, the change in *T*-dependence of $\tau_\beta(T)$ on crossing $T_g$ is one of consequences of the deeper experimental fact that the α- and the JG β-relaxation are coupled or inseparable, and thus $\tau_\beta(T)$ mimics $\tau_\alpha(T)$ in properties. One of the remarkable experimental fact showing that the two relaxations are coupled is the invariance of the ratio of their relaxation times to different combinations of pressure and temperature while keeping one of them constant [98, 99]. The coupling between the two relaxations is the basic prediction of the Coupling Model, wherein the primitive relaxation is the analogue of the JG β-relaxation.

There are only a few papers published in the past reporting neutron scattering on $D_2O$ hydrated proteins that have mentioned the presence of change at $T_g$ [24, 28, 101]. One is the study of hydrated DNA by Sokolov et al. [101] by spectrometer with frequency resolution of ~30 GHz corresponding to ~5 ps in time-scale. These authors integrated the quasi-elastic scattering intensity in the frequency range 60–200 GHz at different temperatures, and rescaled them by the Bose factor. Although the $T_g$ of the hydrated DNA had not been determined either by calorimetry or dielectric relaxation, it can be expected to fall within the range of 180-200 K. By interpolation of the scanty results of the integrated intensity spanning across the possible $T_g$ at 83, 138, 211, 236, 255, and 275 K, Sokolov et al. managed to show its temperature dependence changes at some temperature within 180-200 K. The data had been transformed into the frequency dependence susceptibility spectra $\chi''(\nu)$. We point out that $\chi''(\nu)$ exhibits the nearly constant loss (NCL) at all these temperatures. At 138 and 211 K, the NCL is found at frequencies lower than ~125 GHz. At 236, 255, and 275 K, the NCL is identified with the level of the very flat minimum. When plotted against temperature, these few data points of NCL suggest that the NCL has weak $T$-dependence below the purported $T_g$ within the range of 180-200 K, and changes to a stronger $T$-dependence above it. This property, shared by NCL in many glassformers and hydrated proteins (see also Fig.10), found also in hydrated DNA reaffirms the observation of possible break at the purported $T_g$ is due to NCL from caged dynamics, and its sensitivity to glass transition. Beyond addressing the possible break of integrated intensity at $T_g$, Sokolov et al. analyzed their data at higher temperatures of 297 and 325 K in terms of Mode Coupling Theory (MCT), and suggested there is a dynamic transition at $T$~230 K, which they ascribed to the critical temperature $T_c$ of MCT, but it cannot be the commonly known dynamic transition at $T_d$. The spectrometer they used only can sense motion shorter than ~5 ps (like IN5) or ~20 ps (like IN6) when extended by using 9 Å neutrons. Thus, $T_d$ must be significantly higher than 230 K. This is because a similar system, hydrated tRNA at $h$=0.35-0.50, has $T_d$ near the same temperature from measurement on a spectrometer with frequency resolution of 0.24 GHz, and time scale ~1 ns, a hundredfold better in resolution than in Ref.[102]. Moreover, Cornicchi et al. found $T_d$ for DNA at $h$=0.55 is at ~230-240 K by IN13 with an about 10 times better resolution [22, 103].

Cursory mention of caged motion was made by Khodadadi et al. [40] in their sketch of the protein dynamics. They described the caged dynamics as fast picosecond relaxation, appearing in neutron and light scattering spectra in the frequency range ~100 GHz-1 THz. It is incorrect to restrict manifestation of caged dynamics to such high frequency range. In fact, at temperatures below $T_g$ and in some range above $T_g$, caged dynamics can persist down to much lower frequencies than 100 GHz, and can be observed even by low frequency dielectric and mechanical relaxation as the nearly constant loss. [51, 54, 55]. More importantly, Khodadadi et al. did not mention that caged dynamics can change temperature dependence on crossing $T_g$, which is the thrust of our present paper. From their view of the caged dynamics as fast picosecond relaxation is staying in the frequency range ~100 GHz-1 THz for all temperatures as illustrated in their Fig.11, it cannot cause any change of the $T$-dependence of the mean square displacement from neutron scattering at any temperature including $T_g$.

Another neutron scattering study mentioning change of temperature dependence of the mean square displacement at $T_g$ is on $D_2O$ hydrated lysozyme by Zanotti et al. [24]. In this study the methyl-group rotation in lysozyme has not been taken into account and thus the presence of the change of temperature dependence at $T_g$~150 K is not totally certain. Most proteins fully hydrated by water alone have $T_g$ in the range approximately from 150-180 K, and the complication due to methyl-group rotation contributing in the same temperature range makes it difficult to identify the presence of the break at $T_g$ in the elastic intensity or mean square displacement. This problem is circumvented in solvated proteins using solvents with higher $T_g$ as shown throughout the sections in the above. For proteins hydrated by water alone, full or specific deuteration of the protein or choice of spectrometer with very short time window have to be made to eliminate or remove the methyl group rotation contribution to the observed elastic intensity in order to observe the break at $T_g$. This will be the subject of a paper [104] to follow this one, where the presence of break in $T$-dependence of $<u^2(T)>$ at $T_g$ in several proteins hydrated by water will be proven.

Besides us, Doster had shown in one recent paper the presence of two changes of $T$-dependence of $<u^2(T)>$, one at $T_g$ and another at $T_d$. This he did in Ref.[28] from the proton mean square displacements of $H_2O$ adsorbed to proteins obtained by SPHERE with time resolutions of 2 ns where the protein is C–PC, and by IN6 at 15 ps where the protein is myoglobin. Once more the data show that the dynamic transition temperature $T_d$ depends on the resolution of the spectrometer, but not in the case of the changes of $T$-dependence of $<u^2(T)>$ at $T_g$ found before in conventional glass-formers [47]. The intermediate scattering function $I(Q,t)$ of adsorbed water for myoglobin at $h=0.34$ g/g taken by IN6 shows the decay in two steps within 15 ps. The first step occurring at times less than 0.3 ps was interpreted as associated with fast hydrogen bond fluctuations dubbed the β-process with correlation time nearly independent of $T$ and the wave vector $Q$, while the amplitude increases with $Q$ and $T$. These fast hydrogen bond fluctuations can contribute to $<u^2(T)>$ of caged water molecules, and its $T$-dependence can change at $T_g$ if the fluctuations are sensitive to change in density on vitrification. However, this scenario applies only to hydration $H_2O$, and not when the solvent is glycerol as in Figs.1-3 or glycerol-$D_2O$ in Fig.3. Also this applies not to the $<u^2(t)>$ of water O atoms over a range of temperature ranging from below $T_g$ to some higher temperature but below $T_d$ shown in Fig.10 for hydrated MPB. This is because the fast process seen from $<u^2(t)>$ in Fig.10 extends to 100 ps while the hydrogen bond fluctuations according to Doster manifest at times less than about 0.3 ps. This fast process comes from caged water molecules and has origin common to those found in ordinary glass-formers. It is also present in the case of the hydration water for myoglobin at $h=0.34$ g/g taken by IN6, as evidenced by the plateau of $I(Q,t)$ shown for $T_g \approx 180$ K and 220 K in Ref.[28], and certainly at temperatures below $T_g$ although not presented.

Mode Coupling Theory (MCT) also addressed caged molecular dynamics of conventional glass-formers, and there is a specific prediction on the fast β-process obeying scaling relations. Doster [7, 28] suggested $I(Q,t)$ of adsorbed water for myoglobin at $h=0.34$ g/g taken by IN6 can be explained by the two-steps decay of MCT. Before this can be accepted, the time dependence of $I(Q,t)$ or the susceptibility minimum (obtained after Fourier transformation) and the scaling relations must be tested. Even if successful, it is not clear how MCT can address the change of $T$-dependence of $<u^2(T)>$ at $T_g$ because neither the ideal MCT nor the version that incorporates translation-rotation coupling [105, 106] have predictions for the temperature dependence of the caged dynamics at or near $T_g$.

The crossover of temperature dependence of $<u^2(T)>$ of hydrated and solvated proteins at $T_g$ has been observed before in many small molecular and polymeric glass-formers by neutron scattering [47], dynamic light scattering [50,62], and dielectric relaxation [51]. Hence this property of hydrated protein could likely have the same origin as conventional glass-formers. Proposed before as a possible molecular mechanism responsible for the crossover of temperature dependence of $<u^2(T)>$ at $T_g$ of glass-formers is caged molecular motions, and the sensitivity of the amplitude of the motion to change of specific volume on crossing $T_g$ [62].

The $<u^2(T)>$ having this property corresponds in susceptibility to the loss of molecules moving within cages formed by the anharmonic potential at $T_g$. This loss has no characteristic time and the corresponding $<u^2(T)>$ appears as a scaleless power law or logarithmic function of time (i.e., nearly constant loss in susceptibility), and continues until the cages are dissolved by the onset of the Johari-Goldstein β-relaxation involving the motion of the entire molecule. From this relation between the caged dynamic and the β-relaxation, the change of $<u^2(T)>$ at $T_g$ has been rationalized [62] from the well known and experimentally observed change of $T$-dependence of both the strength and the relaxation time of the β-relaxation on crossing $T_g$ [55,58]. The plenty of evidence for change of $T$-dependence of the nearly constant loss when crossing $T_g$ in common glassformers [47, 50, 51] could be construed to suggest the same interpretation for the solvated proteins. Notwithstanding the suggested origin of $<u^2(T)>$ of hydrated proteins is caged dynamics, more experimental investigation is needed before definitive conclusion can be made. In fact, other possible origin of the observed change of slope of $<u^2(T)>$ at $T_g$ could be vibrational contributions. Some molecular vibrations, including Boson peak intensity and frequency, are also sensitive to temperature change and on crossing $T_g$ [107]. It has been demonstrated that the density of vibrational states of amorphous materials, as well as that of hydrated and solvated proteins [14, 107], undergoes a change on crossing $T_g$. This can affect the elastic intensity of neutron scattering, and the effect reflects in mean square displacement. Therefore, in analyzing only data coming from elastic incoherent neutron scattering, it is difficult to single out or neglect these contributions to $<u^2(T)>$. More experimental work with alternative techniques on hydrated and solvated proteins needs to be done in order to provide an exhaustive picture.

## 3. Conclusion

From the evidences given in the sections above, the following conclusions can be made on the dynamics of solvated proteins probed by elastic or quasielastic neutron scattering, and by molecular dynamics simulations. As a function of increasing temperature and observed from either the protein or the solvent, the mean square displacement $<u^2(T)>$ shows change to a stronger $T$-dependence after crossing temperature close to the glass transition temperature $T_g$, of the solvated protein. In all cases, $T_g$ is independently determined by calorimetry or by dielectric spectroscopy. For solvated proteins, this change of $<u^2(T)>$ at $T_g$ coexists with the usually found dynamic transition of the solvated protein at $T_d$, which is higher than $T_g$. The change of $<u^2(T)>$ is observed near $T_g$, independent of the time window of the spectrometer used for solvated proteins and conventional glass-formers. In contrast, $T_d$ depends on the resolution of the spectrometer because the dynamic transition is caused by entrance inside the time window of the spectrometer of the β-relaxation of the solvent (coupled to the protein) and its most probably relaxation time is a function of temperature.

The change of $<u^2(T)>$ at $T_g$ was found before in many conventional glass-formers of various kinds, and now also in solvated proteins. This supposedly universal dynamic property of glass-formers as well as in hydrated proteins poses as an outstanding problem that challenge for explanation [62]. Possible origin of this general property could be contributions to $<u^2(T)>$ coming from vibrations including the Boson peak by others [107], and caged dynamics by us [47,50,51,62]. Boson peak frequency and intensity [107] show changes on crossing $T_g$, reflecting in the elastic intensity and in a possible crossover of $<u^2(T)>$. Our explanation is the change of amplitude of caged molecular motions on crossing $T_g$, as it has been demonstrated in conventional glass-formers by different techniques (light scattering and dielectric spectroscopy), reporting a crossover in the amplitude of the nearly constant loss. More experimental work is needed to provide a conclusive explanation for the observed change of $<u^2(T)>$ at $T_g$, but from the evidence presented in this paper it is clear that, regardless of its origin, a crossover does exist at $T_g$, it does not depend on the timescale of the spectrometer, and it occurs beside the so-called dynamic transition at $T_d$. Remarkably, molecular dynamic simulations of a hydrated protein have found the change of $<u^2(T)>$ at $T_g$ at times before the dynamic transition at $T_d$.

There are neutron scattering experiments of some hydrated proteins in which the presence of the change of *T*-dependence of $<u^2(T)>$ at $T_g$ is not obvious. This occurs in some hydrated proteins with methyl groups and low $T_g$ that falls in the temperature region where rotation of methyl group contributes strongly to the rise of $<u^2(T)>$ with temperature. Highly hydrated proteins is another case where change of *T*-dependence of $<u^2(T)>$ at $T_g$ cannot be easily observed. A plausible explanation follows from the fact found in conventional glass-formers that the change of *T*-dependence of $<u^2(T)>$ at $T_g$ is small in strong glass-formers with low degree of cooperativity [62], and hydrated water at high hydration level is such the case.

The dynamic transition at $T_d$ can be mistaken as the observed change of *T*-dependence of $<u^2(T)>$ at $T_g$ if one is not aware of the presence of the latter coexisting with the former. This has happened before in the literature, and has led to unnecessary inconsistency in the interpretation of the dynamic transition. The inconsistency is removed after taking into account of the presence the change of *T*-dependence of $<u^2(T)>$ at $T_g$.

**Acknowledgment**

KLN thank CNR-IPCF at Pisa, Italy, and Prof. Pierangelo Rolla and Prof. Mauro Lucchesi of the Dipartimento di Fisica, Università di Pisa, Italy, for hospitality during his stay in the period from March 9th-June 1st, 2011. S.A. acknowledges the support of Galilei Ph.D. School. S.C. thanks J. Wuttke and K. Niss for making available the original published data of mean square displacement of glycerol.